# Institutional Backing and Crypto Volatility: A Hybrid Framework for DeFi Stabilization

Ihlas SOVBETOV[*]

*Department of Economics and Finance, Istanbul Aydin University*

**ABSTRACT**

Decentralized finance (DeFi) lacks centralized oversight, often resulting in heightened volatility. In contrast, centralized finance (CeFi) offers a more stable environment with institutional safeguards. Institutional backing can play a stabilizing role in a hybrid structure (HyFi), enhancing transparency, governance, and market discipline. This study investigates whether HyFi-like cryptocurrencies, those backed by institutions, exhibit lower price risk than fully decentralized counterparts.

Using daily data for 18 major cryptocurrencies from January 2020 to November 2024, we estimate panel EGLS models with fixed, random, and dynamic specifications. Results show that HyFi-like assets consistently experience lower price risk, with this effect intensifying during periods of elevated market volatility. The negative interaction between HyFi status and market-wide volatility confirms their stabilizing role.

Conversely, greater decentralization is strongly associated with increased volatility, particularly during periods of market stress. Robustness checks using quantile regressions and pre-/post-Terra Luna subsamples reinforce these findings, with stronger effects observed in high-volatility quantiles and post-crisis conditions. These results highlight the importance of institutional architecture in enhancing the resilience of digital asset markets.

**Keywords:** Cryptocurrency; Institutional Backing; Price Risk; Volatility; Regulation; Hybrid Framework; Signaling Theory.

---

[*] Corresponding Author. Vice Head of Department, Faculty Member, Asst.Prof.Dr. +902124441428 (Ext.68102) E-mail address: ihlassovbetov@aydin.edu.tr

1. **INTRODUCTION**

The rise of decentralized finance (DeFi) has introduced unprecedented innovation into global financial systems, offering transparent, peer-to-peer alternatives to traditional intermediaries. However, this innovation is accompanied by extreme volatility due to a lack of centralized governance and regulatory oversight (Chen & Bellavitis, 2020; Jensen et al., 2021). Cryptocurrencies remain notoriously volatile, nearly five times that of the S&P 500 (Yae & Tian, 2024). The volatility of Bitcoin prices is almost ten times higher than the volatility of major exchange rates (Baur & Dimpfl, 2021). This level of fluctuation challenges the integration of digital assets into broader financial infrastructure and undermines their potential as reliable stores of value or mediums of exchange.

Price movements are often driven by speculation, social media sentiment, or technical vulnerabilities rather than fundamentals (Sovbetov, 2018; Flori, 2019; Shen et al., 2019; Caferra, 2020; Gupta et al., 2020; Kumar et al., 2021). As DeFi grows in market capitalization and systemic relevance, surpassing $2.6 trillion at its peak in 2021 (Huang et al., 2022), its susceptibility to destabilizing shocks raises urgent questions for financial stability. Amid this, a hybrid approach that blends DeFi's innovation with the credibility and discipline of centralized finance (CeFi) has begun to attract interest. We refer to this emerging model as Hybrid Finance (HyFi).

HyFi posits that the presence of institutional actors (e.g., ETFs, custody banks, regulatory oversight) can signal price reliability, dampening speculation-driven swings and improving market efficiency. Chen et al. (2020) show that institutional investors, especially hedge funds, play a stabilizing role in opaque environments. By substituting for public information providers, institutions enhance price discovery and mitigate mispricing, particularly when market transparency is low.

Institutionally backed crypto assets are more likely to be researched, monitored, and traded by informed investors, leading to more accurate prices and lower volatility (Chen et al., 2020). In the context, tokens with stronger institutional engagement may exhibit lower volatility due to better information flows, surveillance, and risk management (DeVault et al., 2025).

Recent empirical evidence confirms the increasing footprint of institutions in the crypto space. The number of financial institutions holding crypto assets quadrupled from 200 in 2020 to over 846 by 2022 (Huang et al., 2022)(Figure 1). These holdings are concentrated on a small number of highly capitalized assets such as Bitcoin (BTC), Ethereum (ETH), Ripple (XRP), and Binance Coin (BNB) that enjoy better infrastructure, custody solutions, and integration with traditional finance. EY-Parthenon and Coinbase (2025) further reveal that 87% of institutions now hold spot crypto or ETFs (up from 76% in 2024) with BTC (97%), ETH (86%), and XRP (34%) dominating institutional portfolios (see Table 1). These tokens serve as de facto HyFi proxies, characterized by structured governance, public recognition, and deeper integration with traditional finance.

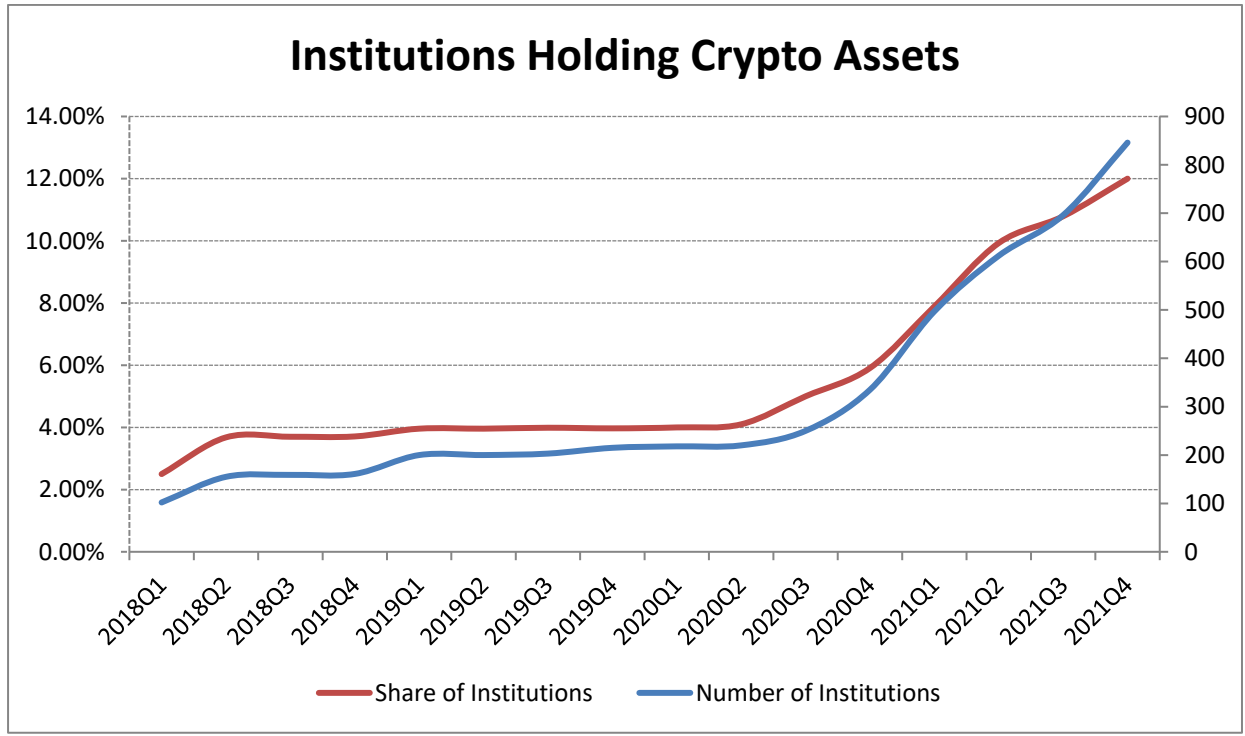

**Figure 1.** Institutions Holding Crypto Assets

**Source:** Huang et al. (2022)

**Table 1.** Most Institutionally Held Cryptocurrencies

| Cryptocurrency | % of Institutional Ownership | Rationale of Institutional Backing |
|---|---|---|
| **Bitcoin (BTC)** | 97% | Widely adopted as a digital store of value; held by major ETFs (e.g., BlackRock, Fidelity); deeply integrated into traditional finance via custody services, derivatives, and spot ETFs; highest institutional penetration. |
| **Ethereum (ETH)** | 86% | Foundation of DeFi applications; enables tokenization and smart contracts; supported by CME futures, ETPs, and institutional staking products; second-most held by institutions. |
| **Ripple (XRP)** | 34% | Tailored for cross-border payments; supported by Ripple Labs with strong banking partnerships; used in enterprise finance; regulatory scrutiny (e.g., SEC litigation) enhanced institutional visibility. |
| **Binance Coin (BNB)** | 24% | Core asset of the Binance ecosystem; used for trading fees, token sales, and platform services; supported by institutional custodians (e.g., Fireblocks); centralized governance and CeFi–DeFi integration position it as a HyFi bridge. |

**Source:** EY-Parthenon and Coinbase (2025)

Despite this momentum, concerns persist. EY-Parthenon and Coinbase (2025) cite an uncertain regulatory environment (52%), volatility (47%), and asset custody (33%) as top barriers to broader adoption (see Figure 2). Meanwhile, among those not yet invested in crypto, lack of valuation fundamentals and fear of financial crimes remain dominant concerns. Notably, 57% of institutions identify regulatory clarity as the key catalyst for future digital asset growth, followed closely by greater institutional adoption. This suggests that trust, structure, and oversight, not decentralization alone, are essential for mainstream integration.

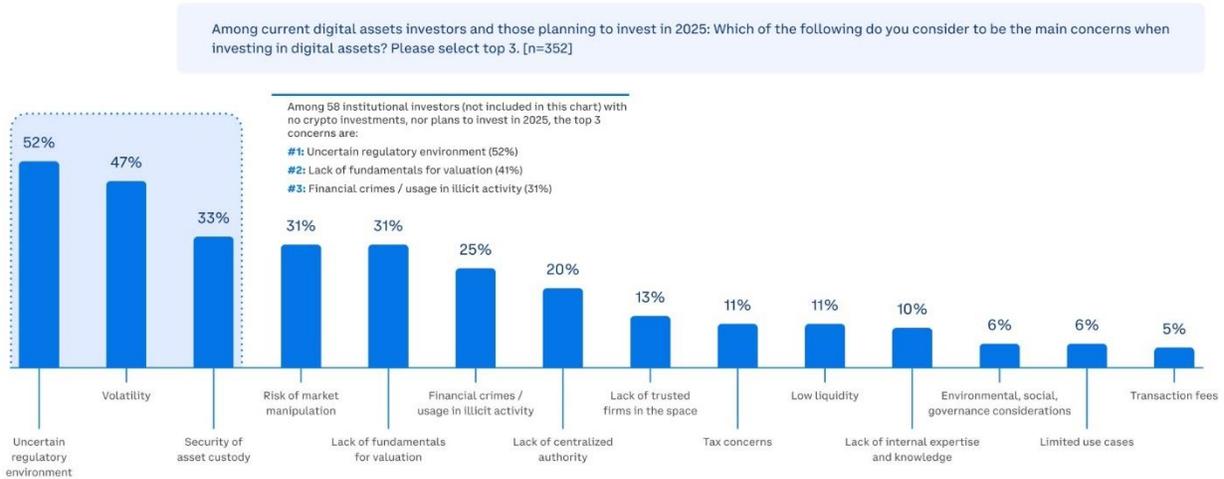

**Figure 2.** Top concerns of institutions investing in crypto assets

**Source:** EY-Parthenon and Coinbase (2025)

Yet, academic research has been slow to empirically evaluate the stabilizing effects of institutional engagement. Most DeFi studies focus on protocol design, governance, or security (Engle et al., 2024; Umar, 2025), but rarely assess how hybrid structures affect market dynamics. The literature on volatility in crypto markets often treats decentralization and institutional presence in isolation (DeVault et al., 2025) rather than in interaction. Our study fills this gap by asking: Do institutionally backed, HyFi-like cryptocurrencies exhibit lower price volatility compared to purely decentralized tokens?

To address this, we construct an empirical framework that incorporates institutional presence, asset-specific factors (decentralization, liquidity, market capitalization), and broader market conditions (volatility, shocks). We focus on whether institutional support dampens volatility directly and whether it moderates crypto assets' response to external shocks. Our study contributes to the growing body of work at the intersection of financial innovation and systemic risk, offering new insights into how hybrid models can promote resilience in the evolving digital asset landscape.

2. Literature Review and Theoretical Foundations

Institutional engagement with digital assets has accelerated in both academia and practice. EY-Parthenon and Coinbase (2025) report that 83% of institutional investors express interest in digital assets provided by regulated financial firms, emphasizing that interoperability between private and public blockchains is a prerequisite for broader institutional adoption. HyFi-like models synthesize these strands by embedding CeFi accountability, disclosure, custody, and risk controls within programmable smart-contract environments to achieve trust, transparency, and resilience.

The growing involvement of centralized financial institutions reinforces this shift toward hybridization. Feldman (2024) documents that more than 65% of major banks in the United States and Europe engage in cryptocurrency-related initiatives, including investment and infrastructure development around Bitcoin (BTC), Ethereum (ETH), Ripple (XRP), and Binance Coin (BNB). Large institutions such as BlackRock, JPMorgan Chase, Goldman Sachs, and BNY Mellon have made substantial financial commitments to these assets and to adjacent services (Andrew, 2024). Public authorities in several jurisdictions have also taken custody of sizeable seized digital-asset holdings, signaling a growing institutional footprint[1]. The United States, China, and the United Kingdom hold significant Bitcoin reserves, underlining an emerging consensus around specific cryptocurrencies as credible stores of value. These developments motivate the working classification of BTC, ETH, XRP, and BNB as HyFi-like assets that are partially embedded in institutional finance and benefit from monitoring, capital commitment, and reputational filtering that are typically absent in purely DeFi tokens.

## 2.1. Conceptual Framework: HyFi as an Interoperability Layer

HyFi is best understood as an interoperability layer that connects the institutional infrastructure of CeFi with the programmability and composability of DeFi. On the CeFi side, regulated entities contribute custody, disclosure, compliance, balance-sheet capital, and reputational screening that raise monitoring quality and create avenues for recourse. On the DeFi side, permissionless smart contracts, automated market making, and modular protocols enable rapid innovation, continuous markets, and efficient settlement. HyFi binds these domains by admitting institutional participation into programmable environments while preserving transparent on-chain execution. Enrollment occurs once through a Digital Universal ID that issues a credential using zero-knowledge proofs; subsequent transactions submit proofs that are verified by a smart-contract verifier, so execution and settlement are non-intermediated after enrollment.

Two mechanisms link this architecture to lower price risk. A signaling channel operates when credible institutional engagement conveys quality and survival prospects to less informed participants, narrowing belief dispersion and reducing the speculative component of order flow. A transaction-cost channel operates when standardized access, custody, audits, and recourse compress search, verification, and enforcement frictions. When these frictions fall, participation widens, liquidity deepens, effective spreads narrow, and price discovery becomes less sensitive to transient order imbalances, which jointly dampens realized volatility. These effects should be strongest when aggregate uncertainty is high, which motivates the interaction between HyFi status and market volatility in Equation (1). This framework maps directly to our empirical design: HyFi-like status captures institutional embeddedness at the asset level; market volatility provides the state variable that governs the strength of both mechanisms; decentralization, liquidity, and size enter as structural controls to avoid confounding institutional embeddedness with network concentration, market depth, or scale.

---

[1] https://bitcointreasuries.net

The diagram in Figure 3 arrays CeFi, HyFi, and DeFi in three columns across the core dimensions (governance, privacy and identity, accessibility, visibility and records, credibility, oversight and recourse, efficiency). For each row, arrows from the CeFi and DeFi cells converge on the corresponding HyFi cell (H1–H7). Color coding indicates HyFi uptake: green denotes full incorporation, gray denotes partial incorporation, and red denotes no incorporation. Beneath each row, two mechanism callouts identify whether that component acts primarily through Signaling Theory (credibility and information effects) or Transaction Cost Theory (reductions in search, verification, monitoring, enforcement, coordination, or settlement costs). The vertical flow terminates in the Outcomes panel, which summarizes the expected effects of HyFi: lower realized volatility and risk, deeper liquidity with narrower spreads, more informative prices, efficient on-chain settlement at lower cost, privacy-preserving compliance, wider institutional participation, and defined oversight with partial recourse.

The conceptual framework is compatible with a concrete architectural instantiation that preserves DeFi's innovation while addressing security, oversight, and governance shortcomings. At its core is a layer-two Digital Universal ID sidechain governed by a Global DAO composed of trusted supranational institutions. Users register identities through a one-time process that generates zero-knowledge proofs; the credential is later verified on the chain at execution, which maintains anonymity while enforcing compliance. By embedding this privacy-preserving identity structure into a smart-contract layer, the HyFi framework enhances security, accountability, and investor protection. The model mitigates risks such as fraud, misuse, and lack of recourse, and it provides a foundation for legally compliant financial products and services. In doing so, HyFi bridges the trust gap that limits DeFi's mainstream adoption and promotes inclusion and stability within a decentralized yet governed digital ecosystem.

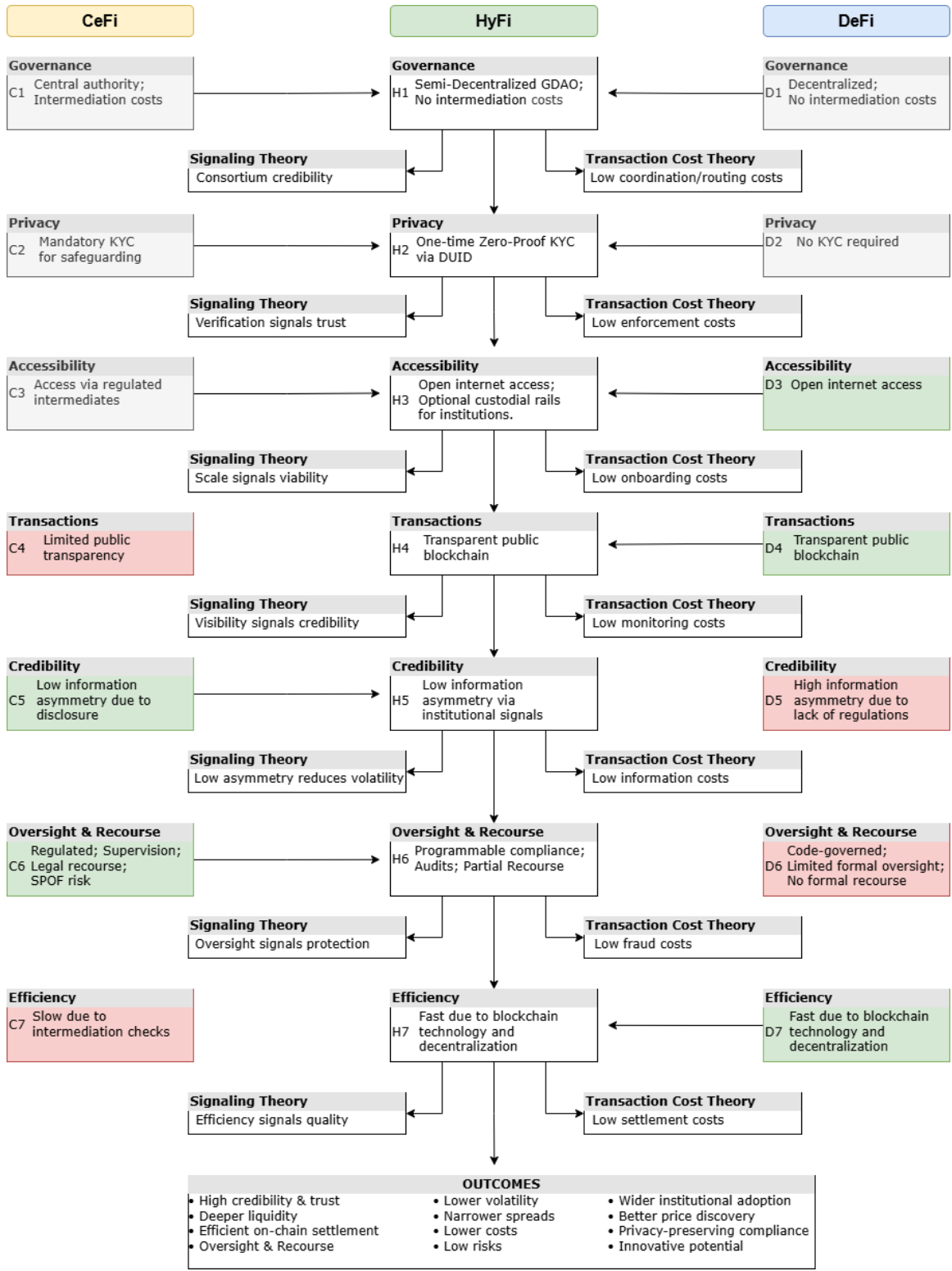

**Figure 3.** HyFi channels linking CeFi and DeFi to market outcomes

Table 2 summarizes comparative features across CeFi, DeFi, and HyFi along dimensions of governance, regulation, trust mechanisms, access, KYC, speed, cost, security, transparency, custody, and innovation.

The table highlights that HyFi retains user custody and on-chain transparency while augmenting DeFi with identity, compliance, and monitoring features that lower risk.

**Table 2.** Key features of CeFi, DeFi, and HyFi

| Feature | CeFi | DeFi | HyFi |
| --- | --- | --- | --- |
| Governance | Centralized (authority) | Decentralized (smart contracts, DAOs) | Semi-Decentralized (smart contracts, GDAO) |
| Regulation | Fully regulated by the authority | No regulation, governed by code | "Code is law", governed by GDAO |
| Trust Mechanism | Trust in the authority | Trust in code | Trust in code and GDAO |
| Accessibility | Requires Bank Accounts | Requires internet access | Requires internet access |
| KYC | Full KYC required | No KYC required | Lightweight KYC (requiring Digital Universal ID) |
| Transaction Speed | Relatively slower, especially cross-border transactions | Fast transactions (blockchain-dependent) | Fast transactions (blockchain-dependent) |
| Cost | Higher fees due to intermediaries | Lower fees (only gas fee) | Lower fees (only gas fee) |
| Security | Highly secure centralized systems, but vulnerable to single points of failure | Highly secure blockchains. High DeFi risks | Highly secure blockchains. Low DeFi risks due to the DUID. |
| Transparency | Limited Transparency | Fully Transparency (on-chain data) | Fully Transparency (on-chain data) |
| Asset Custody | Ensures the security and safety of user assets by the authority | Users retain custody through private keys over their assets | Users retain custody through private keys over their assets, and asset security is enhanced by the DUID |
| Innovation Speed | Slower due to bureaucratic processes | Fast, with permissionless innovation | Fast, with permissionless innovation |

## 2.2. Signaling in Markets with Information Asymmetry

The stabilizing role of institutional backing can be grounded in Signaling Theory, as in Spence (1973). In settings with information asymmetry, institutional investors acting as informed agents signal asset quality, credibility, and risk management practices to less informed market participants. The presence of regulated funds, publicly listed firms, or established financial institutions in a cryptocurrency's ecosystem sends a trust signal that reduces perceived uncertainty. This signal encourages broader participation, improves price discovery, and dampens speculative volatility, particularly in DeFi markets that, while transparent on chain, remain functionally opaque and fragmented because of concentrated governance, pseudonymous participation, and dispersion of liquidity across protocols and chains. These features, coupled with off-chain sentiment and coordination that amplify information asymmetry, are documented in recent work such as Aramonte et al. (2021), Hamrick et al. (2021), and Kim and Kauffman (2024). In this context, HyFi-like cryptocurrencies serve as vehicles of informational clarity, where institutional engagement substitutes for absent regulatory and disclosure norms, aligns expectations, and enhances systemic stability.

### 2.3. Transaction-Cost Economics and Hybrid Intermediation

The stabilizing role of institutional backing is also consistent with Transaction Cost Theory, which predicts that intermediating structures emerge to economize on search, monitoring, and enforcement costs when markets are fragmented and counterparties are opaque (Coase, 1937; Williamson, 2025). In the context of digital assets, HyFi reduces onboarding and discovery costs through standardized access and routing, lowers verification and monitoring costs through privacy-preserving identity, custody, audits, and disclosure, and improves enforcement through programmable compliance and defined avenues for recourse. Compressing these frictions weakens adverse selection, broadens participation, and moderates the volatility of price discovery, with effects that are most pronounced when aggregate uncertainty is high, which motivates the $HyFi_i \times MarketVolatility_t$ term in Equation (1). For an accessible synthesis, see Williamson (2025).

### 2.4. Empirical Implications and Positioning

Evidence from centralized markets complements this argument. Chen et al. (2020) show that when public information deteriorates, hedge funds step in, acquire additional private information, trade more aggressively, and restore market efficiency. Their findings imply that sophisticated investors reduce price risk and information asymmetries, particularly in opaque markets. When institutional participation increases, the adverse effects of diminished public information become statistically insignificant. This substitution effect aligns with microstructure perspectives that link trusted oversight to lower volatility through reduced moral hazard and narrower information gaps, consistent with Goldstein and Yang (2019). In digital assets, this indicates that HyFi-like assets should command lower price risk relative to purely decentralized tokens.

The relevance of this mechanism is heightened by the volatile and off-chain opaque nature of DeFi markets. Reports from the DeFi Rekt database[2] cite cumulative losses exceeding $88 billion arising from speculative trading, smart-contract vulnerabilities, rug pulls, and inadequate oversight. High-profile failures such as the TerraUSD collapse and the Poly Network hack illustrate systemic risks in the absence of enforceable recourse. Zetzsche et al. (2020) warn that a trust-in-code paradigm lacks legal enforceability and leaves users exposed to irrecoverable losses. CeFi incorporates compliance, risk management, and investor protection, and HyFi can replicate these safeguards on-chain through programmable rules and identity.

---

[2] https://de.fi/rekt-database

Our contribution is to move beyond a homogeneous treatment of cryptocurrencies by classifying assets according to their degree of institutional embeddedness and testing whether institutional support, defined as HyFi-like status, systematically reduces price volatility. The empirical strategy controls structural factors such as decentralization, liquidity, and market capitalization. The central prediction is that the interaction between HyFi status and market volatility is negative, consistent with signaling and transaction-cost mechanisms operating more strongly in high-uncertainty states. Such a shift has the potential to expand institutional adoption globally (Conlon et al., 2024). From a risk-pricing perspective, assets with stronger governance should possess a risk premium advantage because the presence of trusted institutions is associated with lower idiosyncratic risk (Duppati et al., 2023). By analogy, HyFi-like cryptocurrencies fall under the oversight of public companies, regulated funds, or consortium banks, which mitigates opportunistic behavior and technical fragility. This perspective parallels agency theory and financial risk theory, where trusted oversight reduces information asymmetries and moral hazard, thereby lowering perceived volatility (De Filippi and Hassan, 2016). KPMG[3] notes that cryptoassets *"present fundamentally new risks"* but also new opportunities to build automated risk controls, and banks entering crypto must *"integrate existing industry frameworks"*. Our classification of assets that straddle DeFi innovation and CeFi safeguards operationalizes these insights.

Crucially, there is limited academic literature on the hybridization of DeFi with CeFi frameworks. While a few studies propose partial forms of integration, such as hybrid credit scoring protocols that combine off-chain data (e.g., credit reports, social media) with on-chain behavior, these efforts remain narrowly focused and raise unresolved regulatory and privacy concerns (Packin & Lev-Aretz, 2024). Most existing research continues to treat cryptocurrencies as a homogeneous asset class and emphasizes macroeconomic or sentiment-driven factors (Sovbetov, 2018; Flori, 2019; Shen et al., 2019; Shen & Wu, 2025), without differentiating tokens based on institutional structure. We address this gap by disaggregating cryptocurrencies according to their degree of institutional embeddedness and empirically evaluating whether institutional support, defined as HyFi-like status, systematically reduces price volatility. Our analysis controls key structural factors such as decentralization, liquidity, and market capitalization. By assessing the stabilizing potential of HyFi-like assets, we offer actionable insights for regulators and investors seeking to integrate digital assets into the broader financial system in a safer and more resilient manner.

### 3. Data & Methodology

#### 3.1. Data & Sample

We classify Bitcoin (BTC), Ethereum (ETH), Ripple (XRP), and Binance Coin (BNB) as proxies for HyFi-like digital assets due to their substantial adoption by institutional investors, as discussed in the preceding conceptual section. Notably, BNB operates as a utility token within Binance's centralized ecosystem while remaining tradable across decentralized networks, making it an archetype of the HyFi structure. These cryptocurrencies have achieved significant recognition and institutional uptake. They embody the core attributes of the HyFi model, namely the integration of decentralized infrastructure with elements of regulatory oversight, institutional trust, and enhanced market stability.

---

[3] https://assets.kpmg.com/content/dam/kpmg/br/pdf/2021/05/plano-bancario-mundo-cripto.pdf

In addition, the study incorporates a balanced sample of other cryptocurrencies that span the major functional categories of the crypto ecosystem:

- **Centralized Exchange Tokens (CEXs):** Cronos (CRO) and OKB.
- **Decentralized Exchange Tokens (DEXs):** Stellar (XLM), Uniswap (UNI), ThorChain (RUNE), Raydium (RAY), Synthetix (SNX), and Gnosis (GNO).
- **Pure DeFi Platforms:** Avalanche (AVAX), ChainLink (LINK), Fantom (FTM), and Polkadot (DOT).
- **Traditional Layer-1 Tokens:** Solana (SOL) and Cardano (ADA), which are integral to the broader cryptocurrency ecosystem.

The selection of these cryptocurrencies is guided by the following criteria:

- **Market Significance:** Inclusion among the top 150 cryptocurrencies by market capitalization during the sampling period, ensuring liquidity and relevance.
- **Functional Diversity:** Purposeful selection across CEX, DEX, DeFi, and traditional chains to capture a wide spectrum of governance and design structures.
- **Institutional Relevance (for HyFi-like assets only):** HyFi proxies are selected based on evidence of institutional participation, regulatory integration, and centralized infrastructure support—characteristics essential to capturing hybrid financial dynamics.
- **Data Availability and Quality:** Daily data availability for open, close, high, low prices, and volume over at least 1,000 trading days to ensure consistency in volatility and risk analysis.
- **Exclusion of Stablecoins and Meme/Fun Coins:** Stablecoins (e.g., USDT, USDC) are excluded due to their fixed pegs, which make them unsuitable for volatility modeling. Similarly, meme-based or entertainment tokens (e.g., DOGE, SHIB) are excluded due to their speculative and often non-fundamental nature, which introduces bias and noise into risk estimations.

This sampling strategy avoids over-representation of speculative assets while ensuring analytical validity across different types of digital assets. We compile daily data (open, high, low, close, trading volume, market capitalization) for 18 major cryptocurrencies from January 1, 2020, through November 24, 2024. Price and market information are obtained from CoinMarketCap[4], which aggregates exchange data worldwide.

Table 3 below presents the list of cryptocurrencies included in the final sample, along with their categories, consensus mechanisms, and data coverage.

---

[4] https://coinmarketcap.com

**Table 3.** Sampled Cryptocurrencies

| NAME | TOKEN | Category | Consensus | Data Range | Obs. |
| --- | --- | --- | --- | --- | --- |
| Bitcoin | BTC | Traditional, HyFi-like | Proof-of-Work (PoW) | 01.01.2020 - 24.11.2024 | 1790 |
| Ethereum | ETH | Traditional, HyFi-like | Proof-of-Stake (PoS) | 01.01.2020 - 24.11.2024 | 1790 |
| Solana | SOL | Traditional | Proof-of-Stake (PoS) | 11.04.2020 - 24.11.2024 | 1689 |
| Ripple | XRP | Traditional, HyFi-like | Proof-of-Consensus (PoC UNL) | 01.01.2020 - 24.11.2024 | 1790 |
| Cardano | ADA | Traditional | Proof-of-Stake (PoS) (Ouroboros) | 01.01.2020 - 24.11.2024 | 1790 |
| Binance | BNB | CEX, HyFi-like | Proof of Staked Authority (PoSA) | 01.01.2020 - 24.11.2024 | 1790 |
| Cronos | CRO | CEX | Proof of Authority (POA) | 01.01.2020 - 24.11.2024 | 1790 |
| OKB | OKB | CEX | Proof-of-Stake (PoS) | 01.01.2020 - 24.11.2024 | 1790 |
| Stellar | XLM | DEX | Proof-of-Agreement (Quorum) | 01.01.2020 - 24.11.2024 | 1790 |
| Uniswap | UNI | DeFi, DEX | Proof-of-Stake (PoS) | 18.09.2020 - 24.11.2024 | 1529 |
| THORChain | RUNE | DeFi, DEX | Proof-of-Stake (PoS) | 01.01.2020 - 24.11.2024 | 1790 |
| Raydium | RAY | DeFi, DEX | Proof-of-Stake (PoS) | 22.02.2021 - 24.11.2024 | 1372 |
| Gnosis | GNO | DeFi, DEX | Delegated Proof-of-Stake (DPoS) | 01.01.2020 - 24.11.2024 | 1790 |
| Synthetix | SNX | DeFi, DEX | Proof-of-Stake (PoS) | 01.01.2020 - 24.11.2024 | 1790 |
| Avalanche | AVAX | DeFi | Snowball Proof-of-Stake (SPoS) | 23.09.2020 - 24.11.2024 | 1524 |
| ChainLink | LINK | DeFi | PoS OCR Off-Chain Reporting | 01.01.2020 - 24.11.2024 | 1790 |
| Fantom | FTM | DeFi | Lachesis Proof-of-Stake (LPoS) | 01.01.2020 - 24.11.2024 | 1790 |
| Polkadot | DOT | DeFi | Nominated Proof-of-Stake (NPoS) | 21.08.2020 - 24.11.2024 | 1557 |

### 3.2. Econometric Setup

The structure of our dataset, which consists of multiple cryptocurrencies (cross-sectional units) observed across time, makes panel data models particularly suitable for empirical analysis. To control unobserved heterogeneity across cryptocurrencies, such as differences in technological maturity, governance structure, or market exposure, the study employs panel Estimated Generalized Least Squares (EGLS)[5] models under both fixed-effects and random-effects specifications. Diagnostic tests indicate significant heteroskedasticity across units (Breusch–Pagan test, $p<0.01$), serial correlation within units (Breusch–Godfrey test, $p<0.01$), and cross-sectional dependence (Pesaran CD test, $p<0.01$). Given these non-spherical error structures, EGLS provides more efficient and consistent estimates by allowing unit-specific error variances, AR(1) disturbances within units, and contemporaneous correlation across units. As robustness, fixed-effects estimates with Driscoll–Kraay standard errors and Common Correlated

---

[5] Estimation is conducted using EGLS, which generalizes the classical panel regression framework by adjusting for heteroskedasticity and potential cross-sectional correlation.

Effects (CCE) models yield qualitatively similar signs and significance, confirming that inference does not hinge on the EGLS specification.

This study investigates whether institutional support, proxied by HyFi-like status, reduces the price risk of cryptocurrencies. The baseline specification is given as:

$$PR_{it} = \alpha_i + \beta_1 HyFi_i + \beta_2(HyFi_i \times MarketVolatility_t) + \sum_{j=1}^{n} \theta_j X_{jit} + \epsilon_{it} \quad (1)$$

where $PR_{it}$ denotes the price risk of cryptocurrency *i* at time *t*, measured using the Parkinson (1980) volatility estimator. The variable $HyFi_i$ is the binary indicator that equals 1 for cryptocurrencies classified as HyFi-like and 0 otherwise. The term $\alpha_i$ captures unobserved individual effects:

- In the Fixed Effects (FE) model, $\alpha_i$ is treated as a fixed constant for each cryptocurrency, controlling for time-invariant heterogeneity.
- In the Random Effects (RE) model, $\alpha_i \sim N(0, \sigma_\alpha^2)$ is treated as a random variable uncorrelated with the regressors.

The vector $X_{it}$ includes control variables that reflect liquidity, size, investor attention, market conditions, decentralization, and exogenous shocks.

To account for liquidity frictions, the model includes the Amihud (2002) illiquidity measure, calculated as the absolute value of daily returns divided by trading volume:

$$Amihud_{it} = \frac{|Return_{it}|}{TradeVolume_{it}}$$

This measure captures the price impact of trading and reflects constraints in liquidity, which can amplify volatility, particularly in thin markets. To enhance interpretability and address scaling issues arising from extremely large trading volumes in the crypto space, the Amihud measure is standardized as a z-score:

$$Z\_Amihud_{it} = \frac{Amihud_{it} - \mu_{i,Amihud}}{\sigma_{i,Amihud}}$$

This transformation ensures that the coefficient reflects the effect of a one-standard-deviation change in illiquidity on price risk.

Cryptocurrency size is captured by the change in the natural logarithm of market capitalization. Investor attention and asset attractiveness are proxied by the logarithm of Google search volume for each cryptocurrency[6]. General market volatility is measured using the realized volatility of a Crypto-50 index, which is constructed following Sovbetov (2018). The index is used to calculate market-wide volatility as:

---

[6] https://trends.google.com/trends/explore

$$RV_t = \sqrt{\frac{1}{L-1} \sum_{k=0}^{L-1} (r_{t-k} - \bar{r}_t)^2}$$

where $r_t = ln(P_t/P_{t-1})$ denotes the daily log return of the top 50 crypto index, and $RV_t$ is calculated as the rolling standard deviation over the past 30 days (with $L = 30$).

To control exogenous market disruptions, the model includes market shock data, represented by the logarithm of funds lost in fraud or scam events, based on the DeFi Rekt database[7].

A central and novel feature of this study is the inclusion of a Decentralization Index, which captures the structural governance characteristics of each cryptocurrency. Unlike prior literature that typically ignores decentralization in empirical models, we incorporate this index to test whether governance dispersion affects price stability. The index is constructed as an equal-weighted average of five Gini coefficients that measure decentralization across distinct domains: (i) network power or stake distribution, (ii) wealth distribution among token holders, (iii) geographical dispersion of validating nodes, (iv) concentration of code contributions on GitHub, and (v) regional distribution of information flow, proxied by tweets.

The dependent variable, price risk, is measured using Parkinson's (1980) volatility estimator, which captures the intraday price range, offering a more information-rich alternative to traditional close-to-close volatility measures. It is calculated as:

$$\sigma_{P,it} = \frac{H_{it} - L_{it}}{C_{it}\sqrt{2Ln2}})$$

where $H_{it}$, $L_{it}$, and $C_{it}$ represent the high, low, and closing prices of asset *i* at time *t*, respectively.

Equation (1) is estimated using both fixed-effects and random-effects panel data models. However, since the fixed-effects estimator may suffer from perfect collinearity when including a time-invariant dummy variable like $HyFi_{it}$, the analysis incorporates an interaction term between the HyFi dummy and market volatility in the fixed-effects specification. This interaction term allows for a more precise examination of how HyFi-like cryptocurrencies behave under varying volatility regimes and isolates their stabilizing role more effectively.

To ensure the robustness of the estimated relationship between institutional support and cryptocurrency price risk, we conduct a series of complementary checks. First, we introduce a lagged dependent variable into the panel fixed-effects specification to account for potential dynamic persistence in volatility. Given the large time dimension of the dataset (T = 1790), the inclusion of a lag term does not raise concerns of Nickell bias and allows for a more flexible representation of short-term risk dynamics[8].

---

[7] https://de.fi/rekt-database

Second, we apply quantile regressions, including Least Absolute Deviations (LAD) at the 10th, 25th, 50th (median), 75th, and 90th percentiles of the price risk distribution. This approach enables us to examine whether the stabilizing effect of HyFi-like status varies across different volatility regimes and offers robustness to outliers and non-normal distributions often observed in crypto markets.

$$Q_T(PR_{it}|X_{it}) = \alpha_\tau + \beta_{1,\tau} HyFi_{it} + \beta_{2,\tau}(HyFi_{it} \times MarketVolatility_t) + \sum_{j=1}^{n} \theta_{j,\tau} X_{it} \qquad (2)$$

where $Q_T(PR_{it}|X_{it})$ denotes the conditional $\tau$-th quantile ($\tau = 0.10, 0.25, 0.50, 0.75, 0.90$) of price risk, and the coefficients $\beta_{1,\tau}$, $\beta_{2,\tau}$, and $\theta_{j,\tau}$ capture the marginal effects at each quantile. Estimation is performed using the Least Absolute Deviations (LAD) method, which is equivalent to median regression when $\tau = 0.50$. The coefficient $\beta_{1,\tau}$ and $\beta_{2,\tau}$ should bring consistent results with equation 1.

Third, we split the sample around the Terra Luna collapse on 7th May 2022, a major shock to decentralized finance ecosystems, and re-estimate the model separately for the pre-crash and post-crash periods. This allows us to assess whether the role of institutional backing differs in relatively stable versus turbulent market environments. Collectively, these robustness checks reinforce the validity of the main findings by accounting for temporal dynamics, distributional asymmetries, and structural breaks.

### 3.3. Decentralization Index

Mainstream literature on decentralization metrics predominantly employs measures such as the Nakamoto coefficient, Shannon entropy, the Herfindahl-Hirschman index (HHI), and the Gini coefficient of power (stakes) (Srinivasan & Lee, 2017; Gencer et al., 2018; Kwon et al., 2019; Lin et al., 2021). While these metrics provide valuable insights, they often focus on single-dimensional aspects of decentralization, potentially overlooking the complex, multifaceted nature of decentralization in blockchain and distributed systems.

To address this limitation, we propose a more comprehensive, multidimensional framework that captures various aspects of decentralization, including network decentralization, wealth decentralization, node decentralization, code decentralization, and information decentralization. This approach is crucial to avoid oversimplifying decentralization and to better account for the diverse layers of decentralization that underpin blockchain systems.

To compute the decentralization index, we use the Gini coefficient, a widely recognized method for measuring inequality, applied across different dimensions of decentralization:

---

[8] Generalized Method of Moments (GMM) estimators are primarily designed for panels with large cross-sectional units (N) and short time dimensions (T). In settings with large T and small N, as in this study, GMM may be inappropriate due to weak instrument problems and overfitting risk.

- **Network decentralization:** This dimension assesses the distribution of power or stakes among validators within the blockchain network. While the Nakamoto coefficient is effective at determining the minimum number of validators required to control 51% of the network, it does not provide insight into how evenly the power is distributed across validators. To address this, we use the Gini coefficient to measure the extent to which power is distributed along a scale from 0 to 1, where a value closer to 1 indicates a higher concentration of power.
- **Wealth decentralization:** This dimension measures the distribution of wealth among token holders. The Gini coefficient is applied to assess how evenly the ownership of tokens is distributed across individuals within the ecosystem.
- **Node decentralization:** This dimension evaluates the distribution of nodes across different geographical regions. Using the Gini coefficient, we analyze the concentration of nodes to determine the extent of decentralization in terms of network infrastructure.
- **Code decentralization:** Code decentralization is measured by the distribution of contributions to the project's codebase. The Gini coefficient is applied to GitHub commits to assess the degree of decentralization in terms of development efforts across contributors.
- **Information decentralization:** This dimension measures the distribution of information flow within the community, as captured by the geographical distribution of tweets related to the cryptocurrency. The Gini coefficient is used to analyze the concentration of information within specific regions.

By combining these multiple dimensions, we construct a multidimensional decentralization index, which reflects the overall decentralization of a cryptocurrency network. This approach provides a more nuanced understanding of decentralization, capturing not only the distribution of power but also the broader ecosystem dynamics that influence the functioning of decentralized networks.

The decentralization coefficients for the sampled cryptocurrencies are summarized in Table 4.

**Table 4.** Decentralization Coefficients

| TOKEN | DECENTRALIZATION COEFFICIENTS | | | | | Equal-Weighted |
|---|---|---|---|---|---|---|
| | **Network** | **Wealth** | **Node** | **Code** | **Information** | |
| BTC | 0.7187 | 0.4675 | 0.3844 | 0.8909 | 0.2849 | 0.5493 |
| ETH | 0.8610 | 0.8148 | 0.4919 | 0.9114 | 0.4572 | 0.7073 |
| SOL | 0.8888 | 0.3965 | 0.4287 | 0.9201 | 0.5930 | 0.6454 |
| XRP | 0.9125 | 0.6385 | 0.3188 | 0.9265 | 0.5392 | 0.6671 |
| ADA | 0.6343 | 0.4351 | 0.4491 | 0.8791 | 0.7028 | 0.6201 |
| BNB | 0.8525 | 0.8700 | 0.4521 | 0.9005 | 0.3863 | 0.6923 |
| CRO | 0.7697 | 0.9464 | 0.4022 | 0.8521 | 0.6814 | 0.7304 |
| OKB | 0.7928 | 0.8799 | 0.4015 | 0.6205 | 0.7695 | 0.6929 |
| XLM | 0.9212 | 0.9105 | 0.3354 | 0.9107 | 0.7772 | 0.7710 |
| UNI | 0.1359 | 0.6930 | 0.4919 | 0.7988 | 0.7044 | 0.5648 |
| RUNE | 0.6702 | 0.4386 | 0.3178 | 0.7948 | 0.7226 | 0.5888 |
| RAY  | 0.1446 | 0.9134 | 0.4287 | 0.7910 | 0.7536 | 0.6062 |
| GNO  | 0.9175 | 0.9550 | 0.3834 | 0.6244 | 0.7720 | 0.7305 |
| AVAX | 0.7589 | 0.5465 | 0.2943 | 0.9208 | 0.6910 | 0.6423 |
| LINK | 0.9064 | 0.6123 | 0.3351 | 0.8340 | 0.6862 | 0.6748 |
| FTM  | 0.9229 | 0.8481 | 0.3659 | 0.7716 | 0.7956 | 0.7408 |
| SNX  | 0.8966 | 0.8248 | 0.5357 | 0.7753 | 0.7786 | 0.7622 |
| DOT  | 0.3229 | 0.3810 | 0.4330 | 0.7998 | 0.6835 | 0.5241 |

**Note:** Data for Network and Nodes gathered from bitnodes.io, etherscan.io, solanacompass.com, livenet.xrpl.org, cexplorer.io, bscscan.com, cronos-pos.org, oklink.com, stellarbeat.io, uniswap.shippooor.xyz, thorchain.net, gnosisscan.io, snowpeer.io, chainlinkecosystem.com, explorer.fantom.network, governance.synthetix.io, polkadot.io, raydium.io. The Gini coefficients for wealth, code, and information are calculated at sovbetov.com/z-gini-economics.php, sovbetov.com/z-gini-commits.php, and sovbetov.com/z-gini-tweets.php respectively.

### 3.4. Data Description and Preliminary Checks

To ensure the validity of our panel estimation strategy, we begin by conducting a series of preliminary diagnostics on the structure and properties of the dataset. Given the panel configuration of 18 cryptocurrencies observed daily over a multi-year period, we first assess the presence of cross-sectional dependence, a common feature in financial data where shocks or volatility in one asset may transmit to others. Results from multiple diagnostic tests, including the Breusch-Pagan LM test, Pesaran Scaled LM test, Bias-Corrected LM test, and the Pesaran CD test, consistently reject the null of cross-sectional independence. These findings confirm significant interdependencies across cryptocurrencies, likely reflecting common exposures to macroeconomic conditions, investor sentiment, or systemic events.

**Table 5.** Unit Root tests

| | CIPS | Truncated CIPS | ADF | Obs. |
|---|---|---|---|---|
| Price Risk | -11.987*** | -6.190*** | - | 24,678 |
| Illiquidity | -16.003*** | -6.190*** | - | 24,660 |
| Size | -20.719*** | -6.190*** | - | 24,660 |
| Decentralization | -2.166** | -2.166*** | - | 24,678 |
| Attractiveness | -4.953*** | -4.759*** | | 32,202 |
| Market Volatility | - | - | 1256.22*** | 32,130 |
| Market Shocks | - | - | 2119.64*** | 32,130 |

**Note:** For cryptocurrency-specific variables, we use Pesaran's CIPS test with a deterministic constant and ADF lags selected by AIC (max lag = 4) to account for cross-sectional dependence. For market-wide variables, cross-sectionally independent ADF tests are applied under the same lag and constant specifications. The null hypothesis assumes a unit root. The ***, **, and * denote significance at the 1%, 5%, and 10% levels, respectively.

To account for this cross-sectional dependence in the analysis of stationarity, we employ the Cross-Sectionally Augmented Im, Pesaran, and Shin (CIPS) unit root test. The CIPS test, which aggregates Cross-Sectionally Augmented Dickey-Fuller (CADF) statistics across units, is designed to provide reliable inference in panels with correlated residuals. The results, reported in Table 5, indicate that all key cryptocurrency-specific variables are stationary at conventional levels of significance. For broader market-level variables such as aggregate volatility and market shocks, which are common across all units, we apply cross-sectionally independent ADF unit root tests. These tests also confirm stationarity, suggesting that no further transformation or differencing of variables is necessary prior to model estimation.

Given the multi-dimensional nature of our independent variables, we examine the potential for multicollinearity using centered Pearson correlation coefficients. While most variables display moderate or low intercorrelation, a notably high association emerges between the decentralization index and log market capitalization, indicating that more centralized tokens tend to attain higher valuations. To mitigate this concern and preserve the conceptual distinctiveness of decentralization, we orthogonalize the decentralization coefficient by regressing it on log market capitalization and extract the residuals, which serve as a *"purified"* decentralization measure for use in all subsequent regressions. This adjustment isolates the independent variation in decentralization, free from confounding size effects.

Table 6 presents the centered correlation matrix for the explanatory variables. Illiquidity, market volatility, and market shocks are moderately correlated with one another, reflecting overlapping market dynamics. However, no pair of variables exceeds a correlation threshold that would suggest problematic multicollinearity.

**Table 6.** Ordinary centered correlation

| Correlation | Illiquidity | Size | Attract. | Decentral. | Mar. Volatility | Mar. Shocks |
|---|---|---|---|---|---|---|
| *Illiquidity* | 1 | | | | | |
| *Size* | 0.1555 | 1 | | | | |
| *Attractiveness* | 0.0547 | -0.0029 | 1 | | | |
| *Decentralization* | 0.0525 | -0.0044 | -0.3455 | 1 | | |
| *Market Volatility* | 0.4267 | -0.1407 | 0.0880 | -0.0022 | 1 | |
| *Market Shocks* | -0.0340 | -0.0179 | 0.0233 | -0.0685 | -0.1105 | 1 |

Descriptive statistics for all variables used in the estimation are provided in Table 7. The average daily price risk, measured via Parkinson's volatility estimator, is 0.063 with a high degree of skewness and kurtosis, indicating the presence of extreme values and fat tails, which are typical in cryptocurrency markets. Illiquidity, expressed as a standardized Amihud measure, is centered around zero by construction but displays significant variation, especially among less liquid tokens. Size dynamics, proxied by the first difference of log market capitalization, show a mean close to zero, capturing daily percentage changes in token size. The distribution is notably skewed and leptokurtic, reflecting large swings in capitalization during volatile periods. Attractiveness exhibits wide dispersion across cryptocurrencies, with several tokens displaying outlier behavior. The decentralization index ranges from 0.52 to 0.86, highlighting considerable variation in governance and structural openness across the sample. The decentralization index (not-orthogonalized) ranges from 0.52 to 0.86, highlighting considerable variation in governance and structural openness across the sample.

**Table 7.** Data descriptive statistics

|  | Price Risk | Size | Illiquidity | Attract. | Decentral. | Mar. Volatility | Mar. Shocks | HyFi |
|---|---|---|---|---|---|---|---|---|
| *Mean* | 0.063 | 0.002 | 0.000 | 11.465 | 0.694 | 0.022 | 5.974 | 0.174 |
| *Median* | 0.049 | 0.001 | -0.283 | 4.000 | 0.699 | 0.015 | 0.000 | 0.000 |
| *Maximum* | 1.436 | 2.168 | 16.231 | 100.000 | 0.865 | 0.286 | 20.897 | 1.000 |
| *Minimum* | 0.003 | -1.937 | -1.053 | 0.000 | 0.523 | 0.000 | 0.000 | 0.000 |
| *Std. Dev.* | 0.054 | 0.063 | 1.000 | 15.921 | 0.079 | 0.025 | 6.727 | 0.379 |
| *Skewness* | 4.284 | 1.721 | 3.701 | 2.211 | -0.212 | 2.652 | 0.374 | 1.724 |
| *Kurtosis* | 47.374 | 119.961 | 30.599 | 8.773 | 2.334 | 17.199 | 1.416 | 3.972 |
| *Obs.* | 30,923 | 30,923 | 30,923 | 30,923 | 30,923 | 30,923 | 30,923 | 30,941 |

Table 8 summarizes the definitions, transformations, data sources, and expected theoretical relationships for each variable included in the econometric model. This structured overview facilitates transparency and aligns each measurement with its empirical role in assessing the determinants of cryptocurrency price risk.

**Table 8.** Data Characteristics

| Variable | Proxy / Transformation | Source | Expected Impact on Price Risk | Reference |
|---|---|---|---|---|
| Price Risk | Parkinson Volatility Estimator | CoinMarketCap | - | - |
| HyFi Status | Dummy variable (0 = non-HyFi, 1 = HyFi-like) | Based on institutional holdings | Negative | - |
| Volatility | Realized Volatility over the past 30 days | CoinMarketCap | Positive | Wang et al. (2023) |
| Size | Log change in market capitalization | CoinMarketCap | Ambiguous | Liu et al. (2022) |
| Illiquidity | Z-score of the Amihud measure | CoinMarketCap | Positive | Liu & Tsyvinski (2021), Bogdan et al. (2023) |
| Market Shocks | Natural log of scam/fraud loss amount | DeFi Rekt database | Positive | Mohamad & Dimitriou (2024) |
| Attractiveness | Natural log of Google search trends | Google Search Trends | Ambiguous | Sovbetov (2018) |
| Decentralization | Gini-based Multi-Dimensional Decentralization (Orthogonalized) | Explained in Section 3.3 | Ambiguous | - |

4. **Results**

**4.1. Core Results**

This section reports the panel EGLS estimates and robustness checks that test whether HyFi-like cryptocurrencies, characterized by institutional support and structured governance, exhibit lower price risk than purely decentralized tokens. Table 9 summarizes results for random-effects (RE) and fixed-effects (FE) specifications, including dynamic variants with a lagged dependent variable to capture persistence in price risk.

Across all specifications, the intercept is positive and highly significant, indicating a baseline level of volatility. Intercepts are modestly lower under FE, which suggests that time-invariant asset heterogeneity absorbs part of this baseline. The Hausman test does not reject the null in the static RE model, which supports the efficiency of RE relative to FE. In the dynamic setting, the Hausman test strongly rejects the null, indicating correlation between regressors and unobserved effects and validating the use of FE in the dynamic specification.

The HyFi-like indicator is negatively associated with price risk in RE and dynamic specifications, consistent with the idea that institutional backing and structured governance reduce information asymmetry and improve monitoring (Chen et al., 2020). Point estimates are economically small but precise, around −0.0091 to −0.0102. Crucially, the FE interaction confirms that HyFi-like assets are substantially less exposed to market-wide turbulence. The coefficient on HyFi × Market Volatility is −0.3422 (p<0.01). This means that for a one-unit increase in standardized market volatility, the increase in price risk is 0.3422 smaller for HyFi-like assets than for non-HyFi assets. In the dynamic FE model, the long-run attenuation equals $\beta_{INT}/(1-\phi) \approx -0.392$ given $\phi = 0.2918$. With standardized volatility, implied short-run slopes are 0.3728 for non-HyFi and 0.0950 for HyFi, and long-run slopes are about 0.526 for non-HyFi and 0.134 for HyFi, which indicates a pronounced flattening rather than a reversal of the volatility–risk relationship. Figure 4 visualizes these patterns, and Figure 5 summarizes the corresponding slopes with approximate confidence intervals.

**Table 9.** Results of the Panel Estimation

|  | PANEL EGLS | | DYNAMIC PANEL | |
| --- | --- | --- | --- | --- |
|  | **RANDOM** | **FIXED** | **RANDOM** | **FIXED** |
| *Intercept* | 0.0512*** (0.0034) | 0.0491*** (0.0012) | 0.0349*** (0.0007) | 0.0350*** (0.0014) |
| *Decentralization* | 0.0740*** (0.0125) | 0.0749*** (0.0084) | 0.0185*** (0.0016) | 0.0308*** (0.0065) |
| *Attractiveness* | 0.0004*** (0.0000) | 0.0004*** (0.0000) | 0.0007*** (0.0002) | 0.0019*** (0.0002) |
| *Size* | -0.1136*** (0.0141) | -0.1139*** (0.0194) | -0.1275** (0.0118) | -0.1263** (0.0205) |
| *Illiquidity* | 0.0347*** (0.0016) | 0.0349*** (0.0009) | 0.0312*** (0.0187) | 0.0319*** (0.0009) |
| *Market Volatility* | 0.4239*** (0.0355) | 0.4797*** (0.0512) | 0.3123*** (0.0187) | 0.3728*** (0.0556) |
| *Market Shocks* | -0.0003 (0.0003) | -0.0003 (0.0001) | 0.0001*** (0.0000) | 0.0001*** (0.0000) |
| *HyFi-like* | -0.0091** (0.0044) | - | -0.0102*** (0.0004) | - |
| *HyFi-like x Market Volatility* | - | -0.3422*** (0.0261) | - | -0.2778*** (0.0285) |
| *Price Risk (-1) ($\phi$)* | - | - | 0.3467*** (0.0095) | 0.2918*** (0.0148) |
| Cross-section Random | S.D: 0.0148 Rho: 0.1885 | - | S.D: 0.0000 Rho: 0.0000 | - |
| Idiosyncratic Random | S.D: 0.0307 Rho: 0.8115 | - | S.D: 0.0273 Rho: 1.0000 | - |
| Hausman Test | 6.4912 [0.3705] | | 3434.2322 [0.0000] | |
| Adj. R-squared | 0.6460 | 0.6782 | 0.7149 | 0.7458 |
| Cross-section Dummy | Random | Fixed | Random | Fixed |
| Period Dummy | No | No | No | No |
| Number of Cryptocurrencies | 18 | 18 | 18 | 18 |
| Total Days | 1,789 | 1,789 | 1,789 | 1,789 |
| Total Observations | 30,923 | 30,923 | 30,923 | 30,923 |

**Note:** White standard errors are given in parentheses. Hausman tests report chi-square statistics with p-values in brackets. ***, **, and * denote significance at the 1%, 5%, and 10% levels, respectively.

The decentralization index exerts a consistently positive and significant effect on price risk across all model variants. A one-unit increase in decentralization is associated with an increase of 0.02 to 0.08 units in price risk, depending on the model. This implies that greater decentralization is associated with increased price volatility. While decentralization is often viewed as a foundational value of blockchain systems, these findings suggest that the lack of centralized oversight may come at a cost in terms of price stability. The results challenge the assumption that decentralization inherently stabilizes markets, instead highlighting its potential fragility. While decentralization enhances censorship resistance and democratic participation, it may also introduce operational inefficiencies and regulatory uncertainty that amplify price fluctuations. The significance and consistency of the decentralization coefficient indicate that governance structure is not a peripheral attribute; it plays a central role in shaping investor expectations and market behavior. This insight is particularly important as regulators and developers weigh trade-offs between decentralization and system stability in the design of next-generation crypto infrastructure.

Attractiveness, proxied by search interest, is positively and significantly associated with price risk. Although the coefficient magnitude is modest (0.0004 to 0.0019), the statistical consistency across models underscores the role of investor attention in amplifying short-term volatility.

The effect of size, measured as a log change in market capitalization, is negative and significant throughout. Coefficients range between -0.11 and -0.13, indicating that larger cryptocurrencies experience lower price risk, likely due to their greater liquidity depth, broader adoption, and reduced sensitivity to speculative pressure. In practical terms, a 1-unit increase in log market capitalization lowers daily price risk by over 0.10 units.

Illiquidity, measured via the standardized Amihud indicator, has a positive and statistically significant effect on price risk in all specifications, with coefficients in the range of 0.03 to 0.035. This aligns with the theoretical expectation that price impact is more pronounced in less liquid tokens, contributing to higher volatility.

Among all covariates, market volatility emerges as the strongest predictor of price risk. Its large and significant coefficients, ranging from 0.31 to 0.48, reflect the endogenous transmission of systemic volatility to individual tokens. Market shocks, measured by the log of scam-related fund losses, are significant only in the dynamic models, suggesting that exogenous disruptions exert delayed effects on volatility.

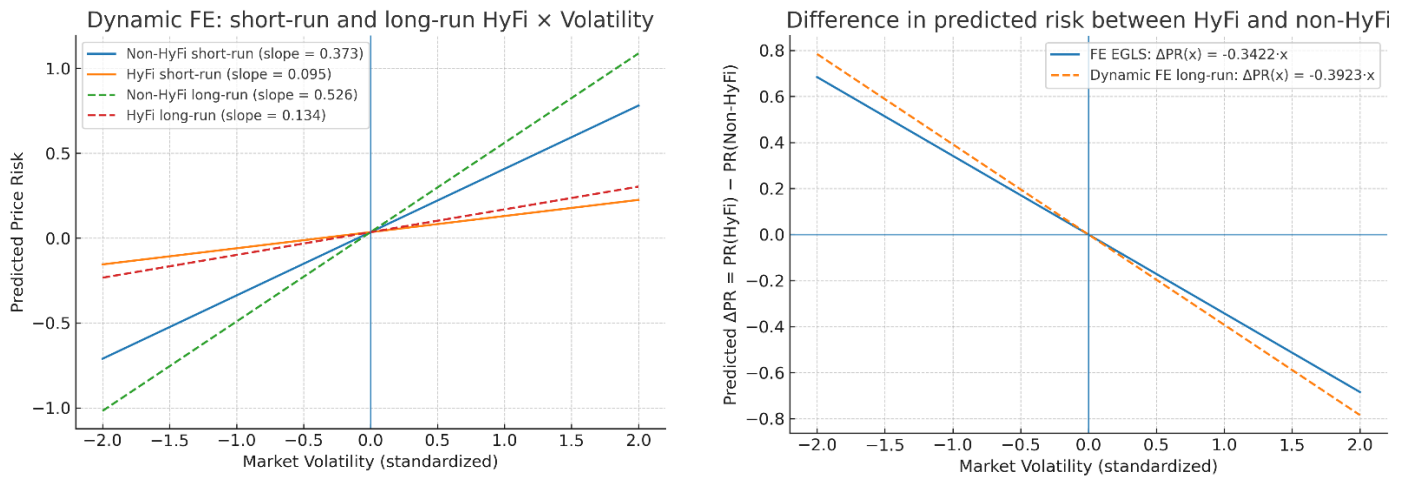

**Panel A:** Predicted price risk vs standardized market volatility for HyFi=0 and HyFi=1. Solid lines are short-run slopes. Dashed lines are long-run slopes computed as $\beta/(1-\phi)$. Here, $\beta = \beta_{MV}$ for non-HyFi and $\beta = \beta_{MV} + \beta_{INT}$ for HyFi; $\phi$ is the coefficient on $PR_{t-1}$.

**Panel B:** Difference in predicted price risk $\Delta PR(x) = PR_{HyFi} - PR_{NonHyFi}$ across volatility. Long–run difference uses $\beta_{INT}/(1-\phi) \cdot x$.

**Figure 4.** Price-risk response to market volatility by HyFi status

**Note:** Market volatility is standardized. HyFi is time invariant and identified through the interaction term. No covariance terms are used in plotting bands; effects are illustrative.

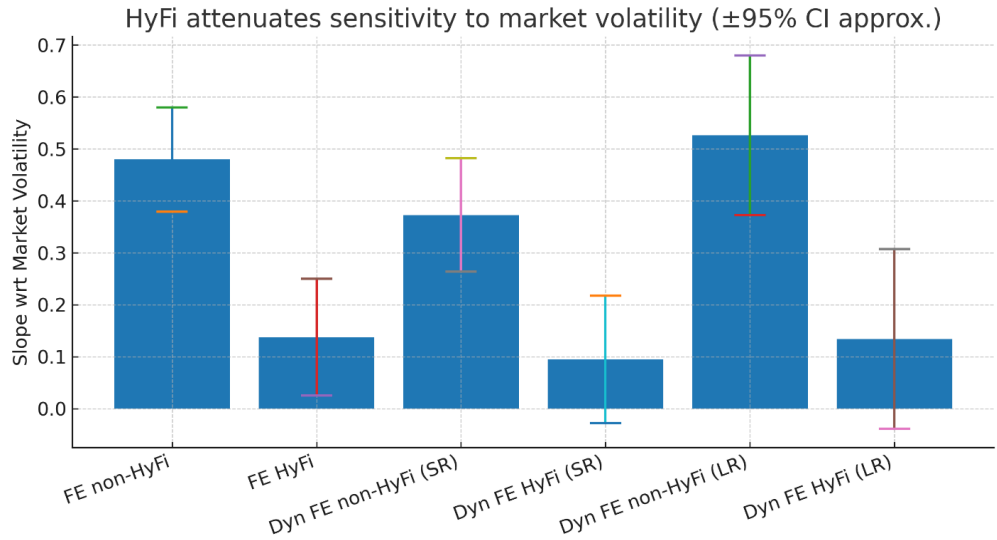

**Figure 5.** Sensitivity of price risk to volatility by HyFi status

**Note:** For FE, the HyFi slope equals $\beta_{MV} + \beta_{INT}$. For dynamic FE, long-run slopes equal the short-run slope divided by $1 - \phi$. Error bars show approximate 95 percent intervals using reported standard errors; covariance between coefficients is not accounted for.

### 4.2. Robustness Checks

To reinforce these findings, we estimate quantile regressions at multiple levels of the price risk distribution (Table 10). Results reveal notable heterogeneity across quantiles. The effects of decentralization, illiquidity, market volatility, and HyFi status intensify in magnitude as one moves toward higher quantiles, indicating their increasing relevance during extreme volatility episodes. For instance, the HyFi dummy coefficient increases (in absolute value) from -0.0074 at the 10[th] percentile to -0.0167 at the 90[th] percentile. This pattern indicates that institutional features not only reduce average volatility but also play a critical role in mitigating tail risks. HyFi-like tokens appear to serve as anchors of relative stability, particularly during stress events that affect the broader crypto ecosystem. The results show that a 1-unit increase in the HyFi dummy is associated with 0.0167 units less volatility in the 90th percentile, signaling substantial crisis resilience.

HyFi effects across quantiles are illustrated in Figure 6. Notably, market shocks are insignificant at lower quantiles but turn significantly negative at the 75[th] and 90[th] percentiles. This implies that, under high-risk conditions, systemic shocks may trigger flight-to-quality behavior, where investors shift toward relatively safer assets, such as HyFi-like tokens. While this does not isolate a specific event, such behavior is consistent with market reactions observed during episodes like the Terra Luna collapse.

Sovbetov, I. (2025). *Computational Economics***Table 10.** Results of the Quantile Regression

| *Tau* | **0.10** | **0.25** | **0.50** | **0.75** | **0.90** |
|---|---|---|---|---|---|
| *Intercept* | 0.0315*** | 0.0385*** | 0.0492*** | 0.0643*** | 0.0822*** |
|  | (0.0003) | (0.0003) | (0.0003) | (0.0005) | (0.0008) |
| *Decentralization* | 0.0014 | 0.0052*** | 0.0207*** | 0.0482*** | 0.0810*** |
|  | (0.0013) | (0.0012) | (0.0015) | (0.0022) | (0.0041) |
| *Attractiveness* | 0.0014 | 0.0005*** | 0.0012*** | 0.0024*** | 0.0041*** |
|  | (0.0013) | (0.0001) | (0.0001) | (0.0001) | (0.0002) |
| *Size* | -0.0618*** | -0.0673*** | -0.0705*** | -0.0778*** | -0.1002*** |
|  | (0.0067) | (0.0050) | (0.0057) | (0.0057) | (0.0086) |
| *Illiquidity* | 0.0259*** | 0.0284*** | 0.0318*** | 0.0370*** | 0.0429*** |
|  | (0.0003) | (0.0003) | (0.0003) | (0.0004) | (0.0006) |
| *Market Volatility* | 0.1724*** | 0.2445*** | 0.3504*** | 0.4977*** | 0.7040*** |
|  | (0.0095) | (0.0088) | (0.0109) | (0.015) | (0.0249) |
| *Market Shocks* | 0.00002 | 0.00002 | -0.00001 | -0.0001*** | -0.0002*** |
|  | (0.00002) | (0.00002) | (0.00002) | (0.0000) | (0.00005) |
| *HyFi-like* | -0.0074*** | -0.0092*** | -0.0118*** | -0.0141*** | -0.0167*** |
|  | (0.0004) | (0.0003) | (0.0003) | (0.0005) | (0.0009) |
| Pseudo R-squared | 0.3074 | 0.3335 | 0.3662 | 0.4071 | 0.4428 |
| Sparsity | 0.0612 | 0.0447 | 0.0504 | 0.0826 | 0.2241 |
| Hall-Sheather BW | 0.0110 | 0.0214 | 0.0310 | 0.0214 | 0.0110 |
| Quantile Dependent Var | 0.0198 | 0.0304 | 0.0486 | 0.0781 | 0.1200 |
| Quasi-LR statistics | 16653.85 | 25761.60 | 29458.67 | 27182.56 | 16039.24 |
| Total Observations | 30,923 | 30,923 | 30,923 | 30,923 | 30,923 |

**Note:** Coefficient covariance matrices are heteroskedasticity-robust (Huber-White sandwich, unweighted). The sparsity function is estimated from residuals using an Epanechnikov kernel with a Hall-Sheather bandwidth (size parameter 0.05) and the Rankit (Cleveland) quantile technique. ***, **, and * denote significance at the 1%, 5%, and 10% levels, respectively.

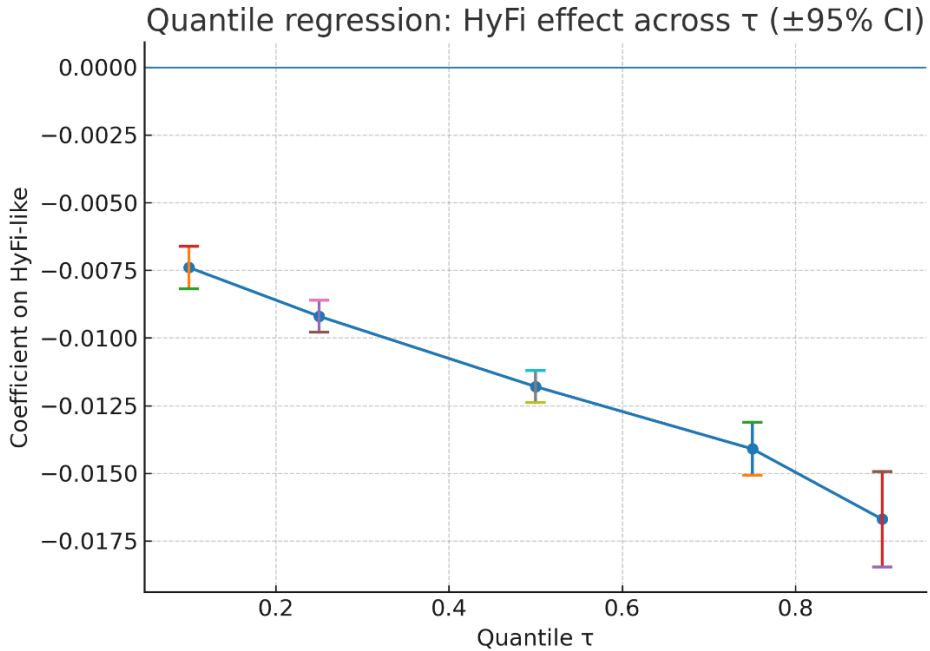

**Figure 6.** Quantile effects of HyFi status on price risk

**Note:** Coefficient paths from quantile regressions at $\tau \in \{0.10, 0.25, 0.50, 0.75, 0.90\}$. Points show estimates for the HyFi-like dummy; vertical bars are ±95% intervals using reported standard errors. Estimates become more negative at higher quantiles, indicating stronger stabilization in the tails.

The role of decentralization also exhibits meaningful variation across quantiles. The coefficient of decentralization increases substantially in higher quantiles (from 0.0014 to 0.081), confirming that highly decentralized tokens are more vulnerable to tail risk. This strengthens the earlier interpretation that governance fragmentation exacerbates instability in extreme market environments. These results not only validate the baseline findings but also show that the stabilizing effect of HyFi-like design and the destabilizing role of decentralization are most pronounced when volatility spikes.

A final robustness check involves a structural break analysis around the Terra Luna collapse of May 2022, one of the largest DeFi failures in history. By re-estimating the model for pre- and post-crash subsamples (Table 11), we assess whether the impact of decentralization and institutional features changed in response to this crisis. The results are striking. Prior to the crash, decentralization was negatively associated with price risk, suggesting a stabilizing role. However, in the post-crash period, this relationship reverses: greater decentralization is now linked to higher volatility, likely reflecting diminished investor confidence in unregulated, fragmented governance systems.

A similar shift is observed in the effect of market shocks. While previously insignificant, market shocks became significantly positive after the Terra collapse, indicating that investors began responding more strongly to systemic risk events. This change reflects heightened market sensitivity and a growing perception that DeFi-related failures can trigger broader contagion effects. Attractiveness also becomes more influential post-crash, while the stabilizing size effect diminishes, and the impacts of illiquidity and market volatility remain relatively stable, though volatility's magnitude slightly declines.

Crucially, the HyFi-like dummy retains its negative and significant association with price risk in both periods, reaffirming the robustness of the central hypothesis. The interaction term with market volatility also remains significantly negative, suggesting that institutional backing continues to buffer systemic shocks even under extreme stress. This consistency reinforces the view that institutional integration not only reduces average volatility but also enhances the long-term resilience of digital assets. To illustrate this attenuation, Figure 7 plots the HyFi × market volatility coefficients from the fixed-effects models before and after the Terra collapse.

**Table 11.** Results of the Panel Estimation

|  | Pre-TerraLuna | | Post-TerraLuna | |
|---|---|---|---|---|
|  | **RANDOM** | **FIXED** | **RANDOM** | **FIXED** |
| *Intercept* | 0.0669*** (0.0046) | 0.0608*** (0.0022) | 0.0445*** (0.0032) | 0.0427*** (0.0012) |
| *Decentralization* | -0.0365*** (0.0068) | -0.0373*** (0.0129) | 0.0351*** (0.0108) | 0.0413* (0.0241) |
| *Attractiveness* | 0.0011*** (0.0002) | 0.0011*** (0.0002) | 0.0063*** (0.0005) | 0.0066*** (0.0007) |
| *Size* | -0.1300*** (0.0159) | -0.1302*** (0.0258) | -0.0841*** (0.0124) | -0.0839*** (0.0238) |
| *Illiquidity* | 0.0341*** (0.0007) | 0.0344*** (0.0011) | 0.0341*** (0.0006) | 0.0341*** (0.0013) |
| *Market Volatility* | 0.4328*** (0.0302) | 0.4855*** (0.0846) | 0.3278*** (0.0172) | 0.3754*** (0.0463) |
| *Market Shocks* | -0.00004 (0.00005) | -0.00005 (0.0001) | 0.0001*** (0.0000) | 0.0001** (0.0000) |
| *HyFi-like* | -0.0323*** (0.0098) | - | -0.0102* (0.0006) | - |
| *HyFi-like x Market Volatility* | - | -0.3191*** (0.0449) | - | -0.2869*** (0.0220) |
| Cross-section Random | S.D: 0.0177 Rho: 0.1772 | - | S.D: 0.0161 Rho: 0.2701 | - |
| Idiosyncratic Random | S.D: 0.0381 Rho: 0.8228 | - | S.D: 0.0207 Rho: 0.7299 | - |
| Hausman Test | 7.9163 [0.2443] | | 0.0000 [1.0000] | |
| Adj. R-squared | 0.6016 | 0.6454 | 0.6916 | 0.7273 |
| Cross-section Dummy | Random | Fixed | Random | Fixed |

| Period Dummy | No | No | No | No |
|---|---|---|---|---|
| Number of Cryptocurrencies | 18 | 18 | 18 | 18 |
| Total Days | 856 | 856 | 933 | 933 |
| Total Observations | 14,129 | 14,129 | 16,794 | 16,794 |

**Note:** White standard errors are given in parentheses. Hausman tests report chi-square statistics with p-values in brackets. ***, **, and * denote significance at the 1%, 5%, and 10% levels, respectively.

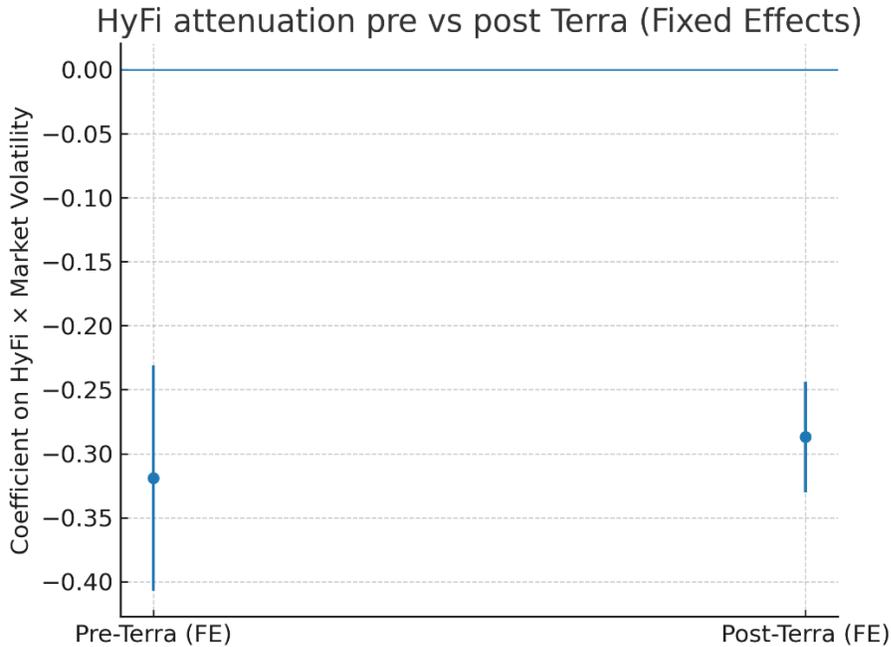

**Figure 7.** HyFi attenuation before and after the Terra collapse (Fixed Effects)

**Note:** Coefficients on HyFi × Market Volatility from FE models estimated pre-Terra and post-Terra with ±95% intervals. Both periods show significant attenuation; magnitudes remain large after Terra.

Collectively, these findings underscore the nuanced interplay between governance structures, market dynamics, and external shocks in shaping cryptocurrency volatility. The consistent negative effect of HyFi-like status on price risk across models suggests that institutional integration plays a meaningful role in stabilizing crypto markets. These insights are relevant not only to investors seeking to manage portfolio risk but also to regulators and policymakers considering the systemic implications of decentralized finance.

## 5. Conclusion

This study investigates whether HyFi-like cryptocurrencies, defined by institutional support and structured governance, exhibit lower price risk compared to their fully decentralized counterparts. Using a high-frequency panel dataset of 18 cryptocurrencies from 1 January 2020 to 24 November 2024, the empirical strategy applies panel EGLS models under both fixed- and random-effects specifications, including dynamic variants that account for persistence in volatility. Robustness is confirmed through quantile regressions across the price risk distribution and structural break analysis around the Terra Luna crash, one of the most disruptive events in DeFi history.

The results consistently reveal that HyFi-like cryptocurrencies are associated with significantly lower price risk. This stabilizing effect becomes even more pronounced during periods of heightened market volatility, suggesting that institutional infrastructure and governance mechanisms play a crucial role in shielding assets from systemic shocks. The robustness of these results across quantile regressions, particularly in the upper tails of the volatility distribution, and during the post-Terra Luna crash period, underscores the resilience conferred by institutional characteristics. These cryptocurrencies appear better insulated against extreme market fluctuations, suggesting that HyFi structures serve as stabilizers in otherwise turbulent crypto markets.

Equally notable is the role of decentralization. The decentralization index is positively and significantly associated with price risk across nearly all models. Contrary to the ideological appeal of decentralization, the empirical evidence suggests that excessive governance dispersion may amplify price volatility. This relationship becomes especially stark during high-volatility episodes and following major market shocks such as the Terra Luna collapse, where decentralized assets exhibited heightened risk. These results highlight the operational and informational fragility that can accompany purely decentralized governance models, particularly in crisis conditions.

Additional determinants such as size, illiquidity, investor attention, market-wide volatility, and external shocks also exert significant effects on price risk, revealing a multifaceted and interdependent volatility structure. Larger cryptocurrencies exhibit lower volatility, while less liquid and highly visible tokens experience greater price swings. Market-wide volatility remains the most dominant driver, but its marginal effect is moderated in the presence of institutional backing. Notably, the impact of these variables intensifies in the tails of risk distribution, indicating nonlinearity in how fundamental features affect extreme outcomes. The insights could be operationalized in regulatory stress testing frameworks or embedded into trading platforms as part of crypto risk-scoring models. For instance, classification based on HyFi-like institutional characteristics could inform portfolio diversification strategies or be integrated into capital adequacy assessments for crypto-exposed financial institutions.

From a policy and implementation standpoint, these findings suggest that HyFi classification could be integrated into regulatory risk assessments, market surveillance systems, or portfolio construction tools. Trading platforms and institutional allocators may benefit from factoring institutional exposure, governance centralization, and volatility interaction effects into crypto asset evaluations.

While the results are robust and the large-T, small-N panel structure mitigates dynamic panel biases such as Nickell bias, limitations remain. These include potential endogeneity from unobserved time-varying factors and partial generalizability due to the focus on mid- to large-cap cryptocurrencies. Nonetheless, the diversity of the sampled tokens, spanning centralized exchanges, DEXs, DeFi platforms, and HyFi assets, ensures relevance across major segments of the crypto ecosystem.

Overall, the study contributes to the emerging literature on institutionalization in digital finance by demonstrating that HyFi-like structures are not only conceptually distinct but empirically associated with greater stability. These insights are highly relevant for regulators, investors, and developers seeking to reduce systemic risk while scaling blockchain-based financial infrastructure. Future research should examine how hybrid governance mechanisms evolve and whether regulatory engagement can further reinforce stability in decentralized asset markets.

**References**

Amihud, Y. (2002). Illiquidity and stock returns: Cross-section and time-series effects. Journal of Financial Markets, 5(1), 31–56. https://doi.org/10.1016/S1386-4181(01)00024-6

Andrew, B. (2024). 6 Biggest Banks Using Ripple (XRP) Products. UseTheBitcoin. Available at: https://usethebitcoin.com/resources/6-biggest-banks-using-ripple-products

Aramonte, S., Huang, W., & Schrimpf, A. (2021). DeFi risks and the decentralisation illusion. BIS Quarterly Review. Bank for International Settlements. http://www.bis.org/publ/qtrpdf/r_qt2112b.pdf

Baur, D.G. & Dimpfl, T. (2021). The volatility of Bitcoin and its role as a medium of exchange and a store of value. Empirical Economics, 61, 2663–2683. https://doi.org/10.1007/s00181-020-01990-5

Bogdan, S., Brmalj, N., & Mujacevic, E. (2023). Impact of liquidity and investors sentiment on herd behavior in cryptocurrency market. International Journal of Financial Studies, 11(3), 97. https://doi.org/10.3390/ijfs11030097

Caferra, R. (2020). Good vibes only: The crypto-optimistic behavior. Journal of Behavioral and Experimental Finance, 28. https://doi.org/10.1016/j.jbef.2020.100407

Chen, Y. & Bellavitis, C. (2020). Blockchain disruption and decentralized finance: The rise of decentralized business models. Journal of Business Venturing Insights, 13. https://doi.org/10.1016/j.jbvi.2019.e00151

Chen, Y., Kelly, B., & Wu, W. (2020). Sophisticated investors and market efficiency: Evidence from a natural experiment. Journal of Financial Economics, 138(2), 316–341. https://doi.org/10.1016/j.jfineco.2020.06.004

Coase, R.H. (1937). The nature of the firm. Economica, 4(16), 386–405. https://doi.org/10.2307/2626876

Conlon, T., Corbet, S., & Oxley, L. (2024). The influence of European MiCa regulation on cryptocurrencies. Global Finance Journal, 63, 101040. https://doi.org/10.1016/j.gfj.2024.101040

De Filippi, P., & Hassan, S. (2016). Blockchain technology as a regulatory technology: From code is law to law is code. First Monday, 21(12). https://doi.org/10.5210/fm.v21i12.7113

Sovbetov, I. (2025). *Computational Economics*

DeVault, L., Turtle, H. J., & Wang, K. (2025). Embracing the future or buying into the bubble: Do sophisticated institutions invest in crypto assets? European Financial Management. Advance online publication. https://doi.org/10.1111/eufm.12542

Duppati, G., Kijkasiwat, P., Hunjra, A. I., & Liew, C. Y. (2023). Do institutional ownership and innovation influence idiosyncratic risk? Global Finance Journal, 56, 100770. https://doi.org/10.1016/j.gfj.2022.100770

EY-Parthenon & Coinbase. (2025). Institutional investor digital assets survey report. Retrieved from EY: https://www.ey.com/content/dam/ey-unified-site/ey-com/en-us/insights/financial-services/documents/ey-growing-enthusiasm-propels-digital-assets-into-the-mainstream.pdf

Engle, R.F., Emambakhsh, T., Manganelli, S., Parisi, L., & Pizzeghello, R. (2024). Estimating systemic risk for non-listed Euro-area banks. Journal of Financial Stability, 75, 101339. https://doi.org/10.1016/j.jfs.2024.101339

Feldman, D. (2024). Which Banks Own Bitcoin: Welcome to Institutional Adoption. Plasbit. Available at: https://plasbit.com/crypto-basic/which-banks-own-bitcoin

Flori, A. (2019). News and subjective beliefs: A Bayesian approach to Bitcoin investments. Research in International Business and Finance, 50, 336-356. https://doi.org/10.1016/j.ribaf.2019.05.007

Gencer, A.E., Basu, S., Eyal, I., Van Renesse, R. and Sirer, E.G. (2018). Decentralization in bitcoin and ethereum networks. International Conference on Financial Cryptography and Data Security, pp. 439-457.

Goldstein, I., & Yang, L. (2019). Good disclosure, bad disclosure. Journal of Financial Economics, 131(1), 118–138. https://doi.org/10.1016/j.jfineco.2018.08.004

Gupta, S., Gupta, S., Mathew, M. & Sama, H.R. (2020). Prioritizing intentions behind investment in cryptocurrency: a fuzzy analytical framework. Journal of Economic Studies, 48(8), pp. 1442-1459. https://doi.org/10.1108/JES-06-2020-0285

Hamrick, J.T., Rouhi, F., Mukherjee, A., Feder, A., Gandal, N., Moore, T., & Vasek, M. (2021). An examination of the cryptocurrency pump-and-dump ecosystem. Information Processing & Management, 58(4), 102506. https://doi.org/10.1016/j.ipm.2021.102506

Huang, X., Lin, J. & Wang, P. (2022). Are institutional investors marching into the crypto market? Economics Letters, 220, https://doi.org/10.1016/j.econlet.2022.110856

Jensen, J., Wachter, V. & Ross, O. (2021). An introduction to decentralized finance (DeFi). Complex Systems Informatics and Modeling Quarterly, 26(2021) pp. 46-54. https://doi.org/10.7250/csimq.2021-26.03

Kim, K., & Kauffman, R. J. (2024). On the effects of information asymmetry in digital currency trading. Electronic Commerce Research and Applications, 64, 101366. https://doi.org/10.1016/j.elerap.2024.101366

Kumar, K.A., Chakraborty, M., & Subramaniam, S. (2021). Does Sentiment Impact Cryptocurrency? Journal of Behavioral Finance, 24(2), 202–218. https://doi.org/10.1080/15427560.2021.1950723

Kwon, Y., Liu, J., Kim, M., Song, D., and Kim, Y. (2019). Impossibility of full decentralization in permissionless blockchains. Proceedings of the 1st ACM Conference on Advances in Financial Technologies, pp. 110-123.

Lin, Q., Li, C., Zhao, X. & Chen, X. (2021). Measuring Decentralization in Bitcoin and Ethereum using Multiple Metrics and Granularities. 2021 IEEE 37th International Conference on Data Engineering Workshops (ICDEW), Chania, Greece, pp. 80-87, https://doi.org/10.1109/ICDEW53142.2021.00022

Liu, Y. & Tsyvinski, A. (2021). Risks and returns of cryptocurrency. Review of Financial Studies, 34(6), 2689–2727. https://doi.org/10.1093/rfs/hhaa113

Liu, Y., Tsyvinski, A., & Wu, X. (2022). Common risk factors in cryptocurrency. Journal of Finance, 77(2), 1133–1177. https://doi.org/10.1111/jofi.13119

Mohamad, A. & Dimitriou, D. (2024). From scam to heist: the impact of cybercrimes on cryptocurrencies. Applied Economics Letters, 1–9. https://doi.org/10.1080/13504851.2024.2384529

Parkinson, M. (1980). The extreme value method for estimating the variance of the rate of return. Journal of Business, 53(1), 61–65. http://dx.doi.org/10.1086/296071

Packin, N.G. & Lev-Aretz, Y. (2024). Decentralized credit scoring: Black box 3.0. American Business Law Journal, 61(2), 91–111. https://doi.org/10.1111/ablj.12240

Spence, M. (1973). Job market signaling. The Quarterly Journal of Economics, 87(3), 355–374. https://doi.org/10.2307/1882010

Shen, D., Urquhart, A. & Wang, P. (2019). Does twitter predict Bitcoin? Economics Letters, 174, pp. 118-122. https://doi.org/10.1016/j.econlet.2018.11.007

Shen, D., & Wu, Y. (2025). The role of Guru investor in Bitcoin: Evidence from Kolmogorov-Arnold Networks. Research in International Business and Finance, 75, 102789. https://doi.org/10.1016/j.ribaf.2025.102789

Sovbetov, Y. (2018). Factors Influencing Cryptocurrency Prices: Evidence from Bitcoin, Ethereum, Dash, Litecoin, and Monero. Journal of Economics and Financial Analysis, 2(2), pp. 1-27

Srinivasan, B.S. & Lee, L. (2017). Quantifying decentralization. Available: https://news.earn.com/quantifying-decentalization-e39db233c28e

Umar, M. (2025). The impact of cyber-attacks on different dimensions of cryptocurrency markets. Technology in Society, 81, 102865. https://doi.org/10.1016/j.techsoc.2025.102865

Wang, J.N., Liu, H.C., Lee, Y.S. & Hsu, Y.T. (2023). FoMO in the Bitcoin market: Revisiting and factors. The Quarterly Review of Economics and Finance, 89, 244-253. https://doi.org/10.1016/j.qref.2023.04.007

Williamson, O.E. (2025). Transaction cost economics. In C. Ménard & M. M. Shirley (Eds.), Handbook of new institutional economics (pp. 47–71). Springer.

Yae, J., & Tian, G. Z. (2024). Volatile safe-haven asset: Evidence from Bitcoin. *Journal of Financial Stability, 73*, 101285. https://doi.org/10.1016/j.jfs.2024.101285

Zetzsche, D.A., Arner, D.W., & Buckley, R.P. (2020). Decentralized Finance. Journal of Financial Regulation, 6(2), 172-203. https://doi.org/10.1093/jfr/fjaa010